\documentclass[aps,prc,superscriptaddress,twocolumn]{revtex4}
\usepackage{graphicx}
\usepackage{epsfig}%
\usepackage{bm}
\usepackage{hyperref}
\usepackage{color}
\usepackage{amsmath}

\begin{document}

\title{Effects of Pauli blocking and in-medium nucleon-nucleon cross sections on the stopping power at low-intermediate energy heavy ion collisions}

\author{Xiang Chen}
\affiliation{China Institute of Atomic Energy, Beijing 102413, China}
\author{Yingxun Zhang}
\email{zhyx@ciae.ac.cn}
\affiliation{China Institute of Atomic Energy, Beijing 102413, China}%
\author{Zhuxia Li}
\affiliation{China Institute of Atomic Energy, Beijing 102413, China}%

\date{\today}
\begin{abstract}
Three typical algorithms of Pauli blocking in the quantum molecular dynamics type models are investigated in the nuclear matter, the nucleus and the heavy ion collisions. The calculations in nuclear matter show that the blocking ratios obtained with the three algorithms are underestimated 13-25\% compared to the analytical values of blocking ratios. For the finite nucleus, the spurious collisions occur around the surface of the nucleus owing to the defects of Pauli blocking algorithms. In the simulations of heavy ion collisions, the uncertainty of stopping power from different Pauli blocking algorithms is less than 5\%. Furthermore, the in-medium effects of nucleon-nucleon ($NN$) cross sections on the nuclear stopping power are discussed. Our results show that the transport models calculations with free $NN$ cross sections result in the stopping power decreasing with the beam energy at the beam energy less than 300 MeV/u. To increase or decrease the values of stopping power, an enhanced or suppressed model dependent in-medium $NN$ cross section is required.
\end{abstract}


\pacs{Valid PACS appear here}
\maketitle

\section{Introduction}
\label{sec:Introduction}

Heavy ion collisions (HICs) provide crucial insights into the features of the nuclear equation of state (EOS) and the in-medium $NN$ cross sections for a wide range of densities, temperatures, and neutron-proton asymmetries. However, the transient state of compressed/expanded nuclear matter during the reaction, such as the pressure and density, can not be directly measured due to spatial-temporal scale of the reaction system which is beyond the capability of measurement. To extract the EOS or in-medium $NN$ cross sections, the transport models which are used to simulate the heavy ion collisions are indispensable.

Many transport codes have been developed to extract the isospin asymmetric nuclear EOS and in-medium $NN$ cross sections~\cite{Aichlin1991PR,AOno1992POTP,SABass1998PPNP,MPapa2005JOCP,BALi2008PR,Zhang2020FOP,Aichlin2020PRC,Bertsch1988PR,JSu2011PRC,ZQFeng2011PRC,
Cozma2013PRC,Nara1999PRC,Nara2016PRC,Niita1995PRC,Ogawa2015PRC,QFLi2005PRC,CMKo1988PRC,Colonna2013PRL,Danie2000NPA}. However, the model dependence of the constraints of symmetry energy~\cite{Tsang04,LWChen05,Tsang09,Rizzo08,Xiao09,ZQFeng10,Xie13,JHong14,TSong15,Cozma17,YYLiu21} and in-medium $NN$ cross sections~\cite{Zhang2007PRC,QFLi2010MPLA,QFLi2011PRC,YJWang2014PRC,Henri2020PRC,Basrak2016PRC} become apparent since the different conclusions could be drawn from the same data. The situations led to the idea of a systematic comparison and evaluation of transport codes under controlled conditions\cite{JXu2016PRC,Zhang2018PRC,AOno2019PRC,Maria21}. Previous studies along this direction were dedicated to the comparison of transport model predictions for Au+Au collisions\cite{JXu2016PRC}, and for benchmarking the treatment of nucleon-nucleon collisions and Pauli blocking\cite{Zhang2018PRC} and $\Delta$ production\cite{AOno2019PRC} in box calculations with cascade mode. The comparisons of the mean field have been done in box calculation with Vlasov mode\cite{Maria21}. 
Currently, the observed differences in the reaction path and corresponding observables are mainly resulted from differences in the initialization of the systems and in the treatment of Pauli blocking effects. The latter, i.e., Pauli blocking, describes the statistical ability to populate the final states in the fermionic system in the gain(loss) term of transport equation for a fermionic system and it is crucial for simulating the low-intermediate energy heavy ion collisions in transport models.

The algorithms of Pauli blocking in Boltzmann-Uehling-Uhlenbeck (BUU) approach and quantum molecular dynamics (QMD) approach are different. In the BUU approach, the Pauli blocking may be improved by increasing the number of test particles to infinity. In the QMD approach, a fixed width of Gaussian wave packet is used to represent a nucleon. It leads to a strong fluctuation, which is important in physics for describing the cluster formation and multifragmentation, but underestimates Pauli blocking effect as in Ref.~\cite{Zhang2018PRC}.
Consequently, one can expect that the successful collision rate of $NN$ collision could be overestimated in the transport codes, and thus, the extracted model dependent in-medium $NN$ cross sections deviate from its real values. Another, owing to the difficulties on the descriptions of Pauli blocking in QMD approach, different Pauli blocking algorithms have been developed in QMD codes~\cite{Zhang2018PRC}.  

The goal of this work is to learn the systematic deviation of Pauli blocking ratio compared to the analytical values and evaluate the uncertainties of nuclear stopping power caused by different Pauli blocking algorithms in the simulations of HICs. All the calculations are performed within the framework of the improved quantum molecular dynamics model (ImQMD)~\cite{Zhang2014PLB,Zhang2015PLB,Zhang2020PRC,Zhang2020FOP}, but the Pauli blocking algorithms are replaced by the three typical algorithms. The paper is organized as follows: In Sec.~\ref{sec:Methods}, we briefly describe the three typical algorithms of Pauli blocking in the market and the in-medium $NN$ cross sections we try to analyzed. In Sec.~\ref{sec:results}, the influences of different Pauli blocking algorithms in the nuclear matter, the finite nucleus and the heavy ion collisions are presented and discussed. Furthermore, a simple discussion on the influence of in-medium $NN$ cross sections on the stopping power is also presented. We do not compare the model calculations to the data for extracting the real values of in-medium $NN$ cross sections in this work, because the algorithms of Pauli blocking still need to be improved. A summary and outlook are given in Sec.~\ref{sec:summary}.

\section{Pauli Blocking and in-medium $NN$ cross sections in ImQMD model}
\label{sec:Methods}
In this section, we only mention the three typical algorithms of Pauli blocking that are directly related to Uehling-Uhlenbeck factor in the transport equation, and the in-medium $NN$ cross sections that we used. There are also some efforts to improve the Pauli blocking in the transport codes by adding additional constraint~\cite{Aichelin1985PRC,Cozma2013PRC,Cozma2018EPJA,QFLi2011PRC}, but we will not touch it in this work. More details about the mean field potential and the treatment of nucleon-nucleon collision in the ImQMD model can be found in Refs.~\cite{Zhang2020FOP,Zhang2014PLB,Zhang2015PLB,Zhang2020PRC}.



\subsection{Algorithms of Pauli blocking}
In the quantum molecular dynamics type models, each nucleon is represented by a Gaussian wave packet,
\begin{equation}
\label{phi1}
\psi_i(\mathbf{r})=\frac{1}{(2\pi\sigma_r^2)^{3/4}}{\rm e}^{-\frac{(\mathbf{r}-\mathbf{r}_{i})^2}{4\sigma_r^2}+{\rm i}\mathbf{r}\cdot \mathbf{p}_{i}/\hbar}, i=1, ..., A\vspace*{0.5mm}
\end{equation}
here, $\sigma_r$ and $\mathbf{r}_{i}$ are the width and centroid of wave packet, respectively. Its Wigner density reads
\begin{equation}
\label{f1qmd}
f_{W,i}(\mathbf{r},\mathbf{p})=\frac{1}{(\pi\hbar)^{3}}e^{-\frac{(\mathbf{r}-\mathbf {r}_{i})^2}{2\sigma_r^2}-\frac{(\mathbf{p}-{\mathbf{p}_{i}})^2}{2\sigma_p^2}},
\end{equation}
where $\sigma_r\sigma_p=\hbar/2$. The Wigner density expresses the probability density of the simultaneous values of $\mathbf{r}$ and $\mathbf{p}$ for $i$th nucleon. When Eq.~(\ref{f1qmd}) is integrated with respect to $\mathbf p$, the correct probability in coordinate space $|\psi_i(\bm{r})|^2$ is given; if we integrate Eq.~(\ref{f1qmd}) with respect to $\mathbf r$, the correct probability in momentum space $|C_i(\mathbf{p})|^2$ can also be verified~\cite{Zhang2020FOP}.

In the treatment of collision of $i+j\to i'+j'$ at certain time step in the code, the positions of particle $i$ and $j$ are kept same before and after collisions, i.e., $\mathbf{r}_i=\mathbf{r}_{i'}$, $\mathbf{r}_j=\mathbf{r}_{j'}$, while the momenta of particle $i$ and $j$ are changed, i.e., $p_i+p_j\to p'_i+p'_j$. Thus, the probability of the final state $p'_i$ occupied by other nucleons, i.e., $P(p'_i)$, can be calculated based on $\sum_{j\ne i} f_{W,j}(\mathbf{r},\mathbf{p})$ in phase space cell around $p'_i$. In the following discussions, we briefly mention the methods on the calculation of $P(p'_i)$, such as PB-Wigner, PB-Husimi, PB-HSP, which are adopted in the different QMD codes.

\subsubsection{PB-Wigner}
For the algorithm of PB-Wigner, the probability of final state $p'_i$ occupied by other particles is expressed as $P_\tau(p'_i)=P_\tau(\mathbf{r}_i, \mathbf{p}'_i)$,
\begin{eqnarray}
&&P_{\tau}(\mathbf{r}_i,\mathbf{p}'_i) = \frac{1}{2/h^3}\sum_{j\in\tau, j\ne i} f_{W,j}(\mathbf{r}_i,\mathbf{p}'_i)\\\nonumber
& = & 4\sum_{j\in\tau(j\neq i)}^{A} \exp \left[-\frac{(\mathbf{r}_{i}-\mathbf{r}_{j})^{2}}{2\sigma_{r}^{2}}\right]\times \exp\left[-\frac{(\mathbf{p}'_{i}-\mathbf{p}_{j})^{2}}{2\sigma_{p}^{2}}\right],
\end{eqnarray}
with $\tau = n$ or $p$. The factor $2/h^3$ results from consideration of the spin in the phase-space cell.

The $P_\tau(p'_i)$ could be larger than 1, because of the fluctuation and the semi-classical transport equation.  If the occupation probability $P_\tau(\mathbf{r}_i,\mathbf{p}'_i)$ is larger than $1$, occupation probability $P_\tau(\mathbf{r}_i,\mathbf{p}'_i)$ is replaced by $\min{(P_\tau(\mathbf{r}_i,\mathbf{p}'_i),1)}$.
This method are used in ImQMD~\cite{Zhang2020FOP,Zhang2014PLB,Zhang2015PLB,Zhang2020PRC}, IQMD-BNU~\cite{JSu2011PRC}, JAM\cite{Nara1999PRC,Nara2016PRC}, JQMD\cite{Niita1995PRC,Ogawa2015PRC}, UrQMD~\cite{QFLi2005PRC,QFLi2011PRC,SABass1998PPNP}. In the ImQMD and UrQMD model, additional criteria is also adopted to enhance the Pauli blocking ratio~\cite{Zhang2014PLB,QFLi2011PRC}, but for convenience, the effect will not be discussed in this paper.


\subsubsection{PB-Husimi}
In the algorithm of PB-Husimi, the probability of final state $p'_i$ occupied by other particles is expressed according to the Husimi function~\cite{Ikeno2020PRC}. The Husimi function is ensured to have a good property as a probability and extra smearing, thus, the distribution is broader than the Wigner function. In detail, the Husimi phase-space distribution function, i.e., $f_H(\mathbf{r},\mathbf{p})$, in the QMD is related to the Wigner phase-space distribution $f_W(\mathbf{r},\mathbf{p})$ as in Ref.~\cite{RFOCon1984PRA,ZXLi1991CPL},
\begin{equation}\label{eq:Husimi1}
f_{H,i}(\mathbf{r},\mathbf{p})=  \int W(\mathbf{r},\mathbf{p} \, \vert \mathbf{r}',\mathbf{p}') f_{W,i}(\mathbf{r}',\mathbf{p}') \mathrm{d} \mathbf{r}' \mathrm{d} \mathbf{p}',
\end{equation}
with
\begin{equation}\label{eq:Husimi2}
W(\mathbf{r},\mathbf{p} \, \vert \mathbf{r}',\mathbf{p}')=\frac{1}{(\pi\hbar)^{3}} \exp \left[ -\frac{(\mathbf{r}'-\mathbf{r})^2}{2\sigma_{r}^{2}}-       \frac{(\mathbf{p}'-\mathbf{p})^2}{2\sigma_{p}^{2}} \right].
\end{equation}
By using Eq.~(\ref{eq:Husimi1}) and Eq.~(\ref{eq:Husimi2}), we obtain the Husimi phase-space distribution function as follows,
\begin{equation}
f_{H,i}(\mathbf{r},\mathbf{p})=\frac{1}{h^{3}} \exp \left[ -\frac{(\mathbf{r}-\mathbf{r}_{i})^2}{4\sigma_{r}^{2}}-       \frac{(\mathbf{p}-\mathbf{p}_{i})^2}{4\sigma_{p}^{2}} \right].
\end{equation}
Thus, the occupation probability at $\mathbf{p}=\mathbf{p}'_i$ and $\mathbf{r}=\mathbf{r}_i$ can be obtained as $P_\tau(p'_i)=P_{\tau}(\mathbf{r}_i,\mathbf{p}'_i)$,
\begin{eqnarray}
\label{Phusimi}
&&P_{\tau}(\mathbf{r}_i,\mathbf{p}'_i) = \frac{1}{2/h^3}\sum_{j\in\tau, j\ne i} f_{H,j}(\mathbf{r}_i,\mathbf{p}'_i)\\\nonumber
& = & \frac{1}{2}\sum_{j\in\tau(j\neq i)}^{A} \exp \left[-\frac{(\mathbf{r}_{i}-\mathbf{r}_{j})^{2}}{4\sigma_{r}^{2}}\right]\times \exp\left[-\frac{(\mathbf{p}'_{i}-\mathbf{p}_{j})^{2}}{4\sigma_{p}^{2}}\right].
\end{eqnarray}
Similarly, if the occupation probability $P_{\tau}(\mathbf{r}_i,\mathbf{p}'_i)$ is larger than $1$, $P_{\tau}(\mathbf{r}_i,\mathbf{p}'_i)=\min{(P_{\tau}(\mathbf{r}_i,\mathbf{p}'_i),1)}$.


\subsubsection{PB-HSP}
In the algorithm of PB-HSP, the occupation probability $P_{\tau}(\mathbf{r}_i,\mathbf{p}'_i)$ is calculated as $P_\tau(p'_i)=P_{\tau}(\mathbf{r}_i,\mathbf{p}'_i)$,
\begin{equation}
  P_{\tau}(\mathbf{r}_i,\mathbf{p}'_i)=\sum_{j\in\tau(j\neq i)}^{A} (O_{ij}^{(x)}/\frac{4}{3}\pi R_{x}^{3}) (O_{ij}^{(p)}/\frac{4}{3}\pi R_{p}^{3}),
\end{equation}
where, $O_{ij}^{(x)}(O_{ij}^{(p)})$ is the volume of the overlap region of spheres with the radius $R_{x}(R_{p})$  of nucleons $i$ and $j$ in coordinate (momentum) space. This method has been adopted in QMD~\cite{Aichelin1985PRC}, LQMD~\cite{ZQFeng2011PRC} and TuQMD (the additional constraints named as surface correction is also adopted in QMD and TuQMD)~\cite{Cozma2013PRC,Cozma2018EPJA}. The $O_{ij}^{(x)}$ is calculated by
\begin{equation}
 O_{ij}^{(x)}=\left\{
 \begin{array}{ll}
    0 \, , & L_{ij}^{(x)}\geq 2R_{x}\\
    \frac{4}{3}\pi R_{x}^{3}-\pi L_{ij}^{(x)} \left[ R_{x}^{2}-\frac{1}{3}(\frac{L_{ij}^{(x)}}{2})^2 \right] , & L_{ij}^{(x)}< 2R_{x}
  \end{array}
\right.
\end{equation}
where $L_{ij}^{(x)}=|\mathbf{r}_i-\mathbf{r}_j|$. 
By replacing $L_{ij}^{(x)}$ to $L_{ij}^{(p)}=|\mathbf{p}'_i-\mathbf{p}_j|$ and $R_x$ to $R_p$, one can obtain $O_{ik}^{(p)}$.  
Here, we take $R_{x}$=3.367 fm , $R_{p}$=112.5 MeV/$c$ as in Ref.~\cite{Zhang2018PRC}. This method corresponds to the uniform phase-space density distribution in coordinate and momentum space, i.e., one nucleon occupies the phase space cell with the size $\frac{4}{3}\pi R_{x}^{3}\cdot \frac{4}{3}\pi R_{p}^{3}=\frac{h^{3}}{2}$. Similarly, if the occupation probability $P_\tau(\mathbf{r}_i,\mathbf{p}'_i)$ is larger than $1$, occupation probability should be replaced by $\min (P_\tau(\mathbf{r}_i,\mathbf{p}'_i),1)$.

\subsection{In-medium $NN$ cross sections}

In the ImQMD, the isospin dependent $NN$ cross sections and the differential cross sections in free space are taken from Ref.~\cite{Cugnon1996NIM}. The in-medium $NN$ cross sections in the ImQMD model is named as $\sigma^{med}_{QMD}$, and $\sigma^{med}_{QMD}=R*\sigma^{free}$, where the medium correction factor $R=(1+\eta(E_{beam})\rho/\rho_0)$. Here, we want to stress the model dependent $\sigma^{med}_{QMD}$ is not exactly same as in true $\sigma^{med}$, and we will show it in the nuclear matter calculations in Sec.~\ref{BOX}. 

\section{Results and discussion }
\label{sec:results}
In this section, we will first evaluate the algorithms of Pauli blocking in nuclear matter in cascade mode. Then, the effects of Pauli blocking on the finite nucleus and on heavy ion collision observable, i.e., stopping power, are presented and discussed. The uncertainties of different Pauli blocking algorithms on the stopping power are obtained.

\subsection{Pauli blocking in nuclear matter}
\label{BOX}


During the heavy ion collisions, various effects interplay and propagate during the reaction process, thus, it is hard to evaluate the Pauli blocking only in the heavy ion collisions. To disentangle the interplay between mean field and $NN$ collisions, we first analyze the nucleon number distribution of final state $\mathbf p'_i$ (i.e., $dN/dp'$) and occupation probability of final state $\mathbf p'_i$, i.e., $P_\tau(p'_i)$. The calculations are performed in the cascade mode, i.e., without mean-field potential, in the nuclear matter. 

To simulate the nuclear matter, a box with imposed periodic boundary conditions is adopted. The periodic boundary conditions are considered as same as in Ref.~\cite{Zhang2018PRC}, i.e., the dimensions of the cubic box are $L_{\alpha}$ = 20 fm, $\alpha\equiv x,y,z$, and the position of the center of box is ($L_x/2, L_y/2, L_z/2$). The box is initialized with a finite temperature in uniform nuclear matter. The density is $\rho = 0.16$ fm$^{-3}$ and isospin asymmetry equals zero, which corresponds to 1280 nucleons (640 neutrons + 640 protons) in the box. In coordinate space, the positions of nucleon are initialized randomly from $0$ to $L_{\alpha}$. In momentum space, the momenta of nucleon are initialized according to the Fermi-Dirac distribution, $f=1/\{1+$exp$[(\epsilon - \mu)/T]\}$, with $\epsilon = p^{2}/2m$, nucleon mass $m$ = 939 MeV/$c^2$, chemical potential $\mu$ = 36.84 MeV, and temperature $T$ = 5 MeV. The $NN$ collision cross sections are set as 40 mb.
\begin{figure}[htbp]
\centering
\includegraphics[angle=0,scale=0.337]{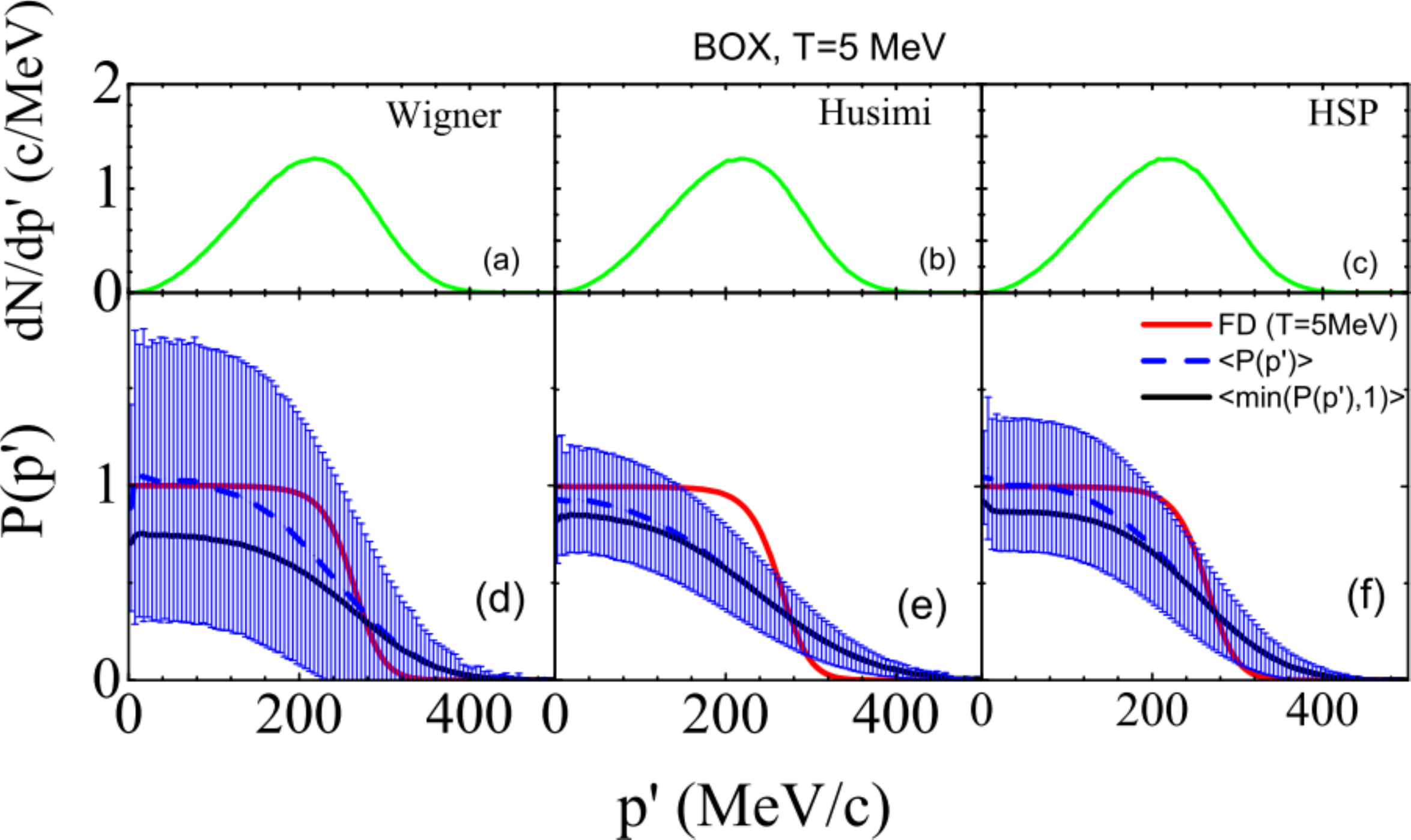}
\setlength{\abovecaptionskip}{0pt}
\vspace{0.9em}
\caption{(Color online) Panels (a), (b), and (c) are the momentum distribution of final state of nucleon-nucleon collisions at t=1 fm/$c$. Panels (d), (e), and (f) are occupation probability for different Pauli blocking algorithms, i.e., PB-Wigner, PB-Husimi, PB-HSP. Red lines are for the analytical values of occupation probability (see text for more details).}
\setlength{\belowcaptionskip}{0pt}
\label{fig:box-Occ-Prob}
\end{figure}

In the upper panels of Fig.~\ref{fig:box-Occ-Prob}, we present the $dN/dp'$ after the evolution of one time step, i.e., at $t=1$ fm/$c$, to understand how many nucleons participate in the $NN$ collisions at certain momentum. Panels (a), (b), and (c) are for the cases of PB-Wigner, PB-Husimi, and PB-HSP, respectively. 10,000 events are performed at $T$ = 5 MeV. The $dN/dp'$ of PB-Wigner, PB-Husimi, and PB-HSP are almost same, because the momentums of nucleons in initial state are sampled within the same Fermi-Dirac distribution and only one time step is considered. $dN/dp'$ increases with momentum and reaches maximum around 240 MeV/$c$ and then decreases.

Fig.~\ref{fig:box-Occ-Prob} (d), (e), and (f) show the occupation probability for the Pauli blocking algorithms of PB-Wigner, PB-Husimi, PB-HSP, respectively. The mean values of the occupation probability, i.e., $\langle P(p')\rangle$, and their standard deviation, i.e. $\langle (P(p')-\langle P(p')\rangle)^2 \rangle^{1/2}$, are shown as the blue curves and the blue error bars. The actual averaged occupation probabilities used in the ImQMD calculations, i.e., $\langle$min$(P(p'),1)\rangle$, are shown as the black curves. The actual occupation probability $\langle$min$(P(p'),1) \rangle$ is always slightly lower than $\langle P(p')\rangle$ due to the truncation of $P(p')$ by using min$(P(p'),1)$. Red lines are the analytical values of occupation probability, i.e., the Fermi-Dirac occupation probability. Among these three Pauli blocking algorithms, the PB-Husimi has a smallest standard deviations due to the large width in Eq.~(\ref{Phusimi}). Furthermore, all those algorithms used in the QMD codes deviate from the theoretical values as discussed in Ref.~\cite{Zhang2018PRC}. It underestimates the Pauli blocking probability at lower momentum region and overestimates the blocking probability at high momentum region due to the strong fluctuation.

\begin{figure}[htbp]
\centering
\includegraphics[angle=0,scale=0.314]{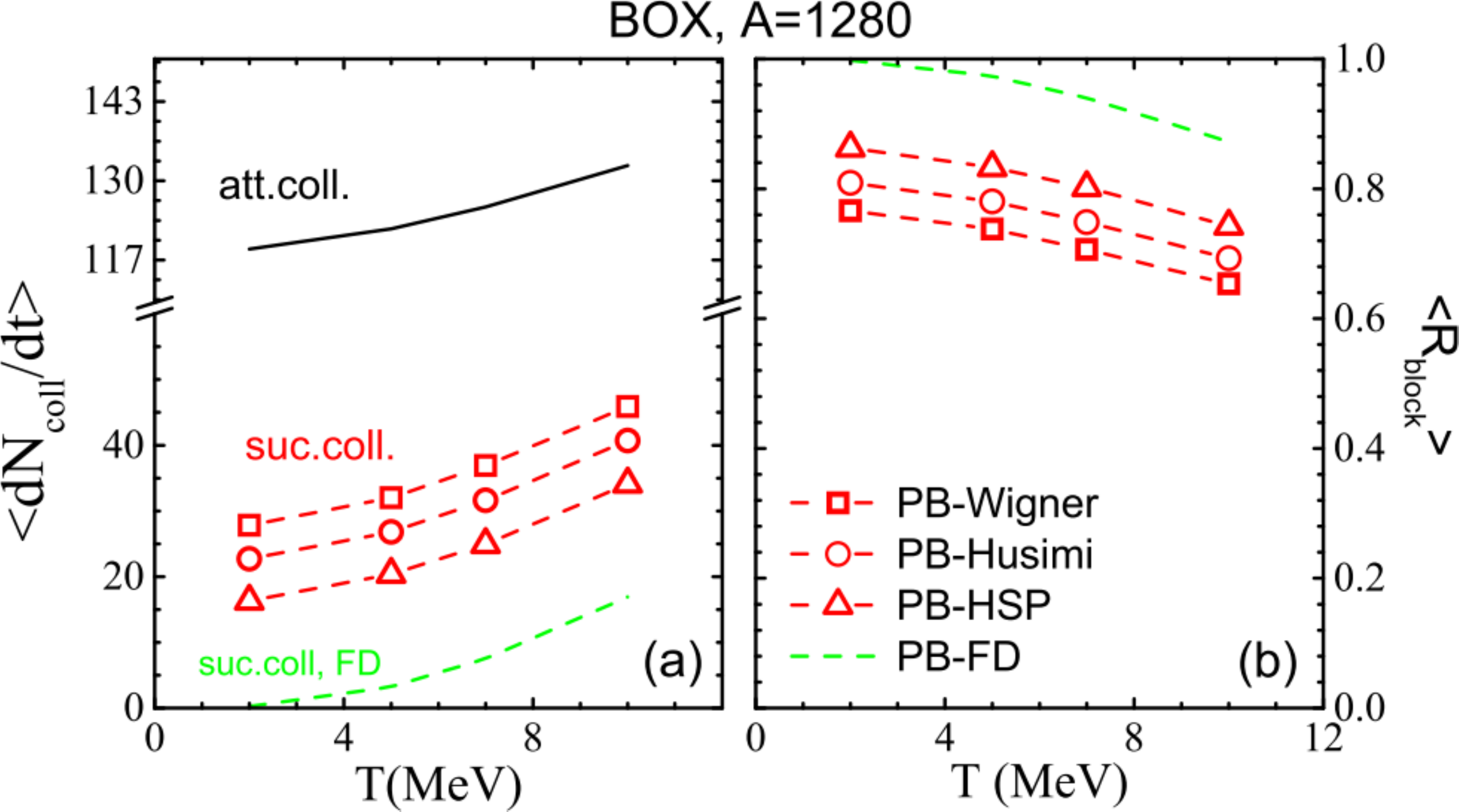}
\setlength{\abovecaptionskip}{0pt}
\vspace{0.0em}
\caption{(Color online) Panel (a): Attempted collision rate $\langle\frac{dN^{att.}_{coll.}}{dt}\rangle$ and successful collision rate $\langle\frac{dN^{suc.}_{coll.}}{dt}\rangle$ as functions of temperature. The red lines with different symbols are from different Pauli blocking algorithms. Panel (b): The averaged blocking ratio $\langle R_{block}\rangle$ for different Pauli blocking (see text for more details).}
\setlength{\belowcaptionskip}{0pt}
\label{fig:box-pbk-anal}
\end{figure}

To quantitatively evaluate the algorithms of Pauli blocking, we present the attempted collision rate $\frac{dN^{att.}_{coll}}{dt}$, successful collision rate $\frac{dN^{suc.}_{coll.}}{dt}$, and Pauli blocking ratio $R_{block}$ defined as,
\begin{equation}\label{Rblk}
R_{block}=1-\frac{dN^{suc.}_{coll.}}{dt}/\frac{dN^{att.}_{coll.}}{dt},
\end{equation}
as a function of temperature in Fig.~\ref{fig:box-pbk-anal} (a) and (b), respectively. The temperature $T$ is from 2 MeV to 10 MeV, which corresponds to the low-intermediate energy heavy ion collisions. The black line is the result for attempted collisions rate which is calculated as in Ref.~\cite{Zhang2018PRC}, i.e.,
\begin{equation}
\label{collrate}
\langle \frac{dN^{att.}_{coll.}}{dt}\rangle =\frac{1}{2}A\rho\langle v_{rel}\sigma^{med}\rangle.
\end{equation}
Here, $A$ is the nucleon number, $\rho$ is the density, $v_{rel}$ is the relative velocity between two colliding nucleons, $\sigma^{med}$ is the in-medium $NN$ cross section and $\sigma^{med}=40$ mb. Eq.(\ref{collrate}) corresponds to the case of no Pauli blocking. The bracket $\langle \cdot \rangle$ means the averaged values over the time interval 60-140 fm/$c$ and from 200 events. As illustrated in Fig.~\ref{fig:box-pbk-anal} (a), the averaged attempted collision rate (black line) increases with the temperature increasing.


The red lines with different symbols in Fig.~\ref{fig:box-pbk-anal} (a) are the averaged successful collision rates obtained with PB-Wigner (square), PB-Husimi (circle), PB-HSP (triangle). 
The green line represents the result of the Pauli blocker which is fixed to the initialized Fermi-Dirac distribution for given temperatures, i.e., PB-FD. It is used for evaluating how well the Pauli blocking algorithms is in the nuclear matter. Similar as the results in Ref.~\cite{Zhang2018PRC}, the averaged successful collision rates from PB-Wigner, PB-Husimi and PB-HSP are larger than the analytical successful collision rates from PB-FD. Among them, the results from PB-HSP are more closer to the PB-FD than others. 

In Fig.~\ref{fig:box-pbk-anal} (b), we plot the averaged blocking ratio $\langle R_{block}\rangle$ as a function of $T$. 
The values of $\langle R_{block}\rangle$ from different Pauli blocking algorithms decrease with temperature increasing. However, the $\langle R_{block}\rangle$ obtained with PB-Wigner, PB-Husimi, and PB-HSP are smaller about 13-25\% than the analytical values, i.e., $\langle R_{block}\rangle$ obtained with PB-FD. It means that the different Pauli blocking algorithms used in the transport codes overestimate the successful collision rate. Thus, to get the same successful collision rate from the algorithms of Pauli blocking adopted in QMD and from the analytical Pauli blocking, one can image that the $\sigma^{med}_{QMD}$ is smaller than its true values, i.e., $\sigma^{med}_{QMD}<\sigma^{med}$. For example, the values of $\xi=\sigma^{med}_{QMD}/\sigma^{med}\approx 37-50\%$ for the selected three Pauli blocking algorithms at $T=10$ MeV.


\subsection{Pauli blocking in a finite nucleus}
\label{Nucleus}
One should notice that the Pauli blocking effect in the nuclear matter is not exactly the same as that in the finite size system due to the boundary effect. Thus, we further check it in a finite nucleus where the mean-field potential is also considered for binding the nucleons together. In the following calculations, the interaction parameter set of SkM* is adopted.



\begin{figure}[htbp]
\centering
\includegraphics[angle=0,scale=0.32]{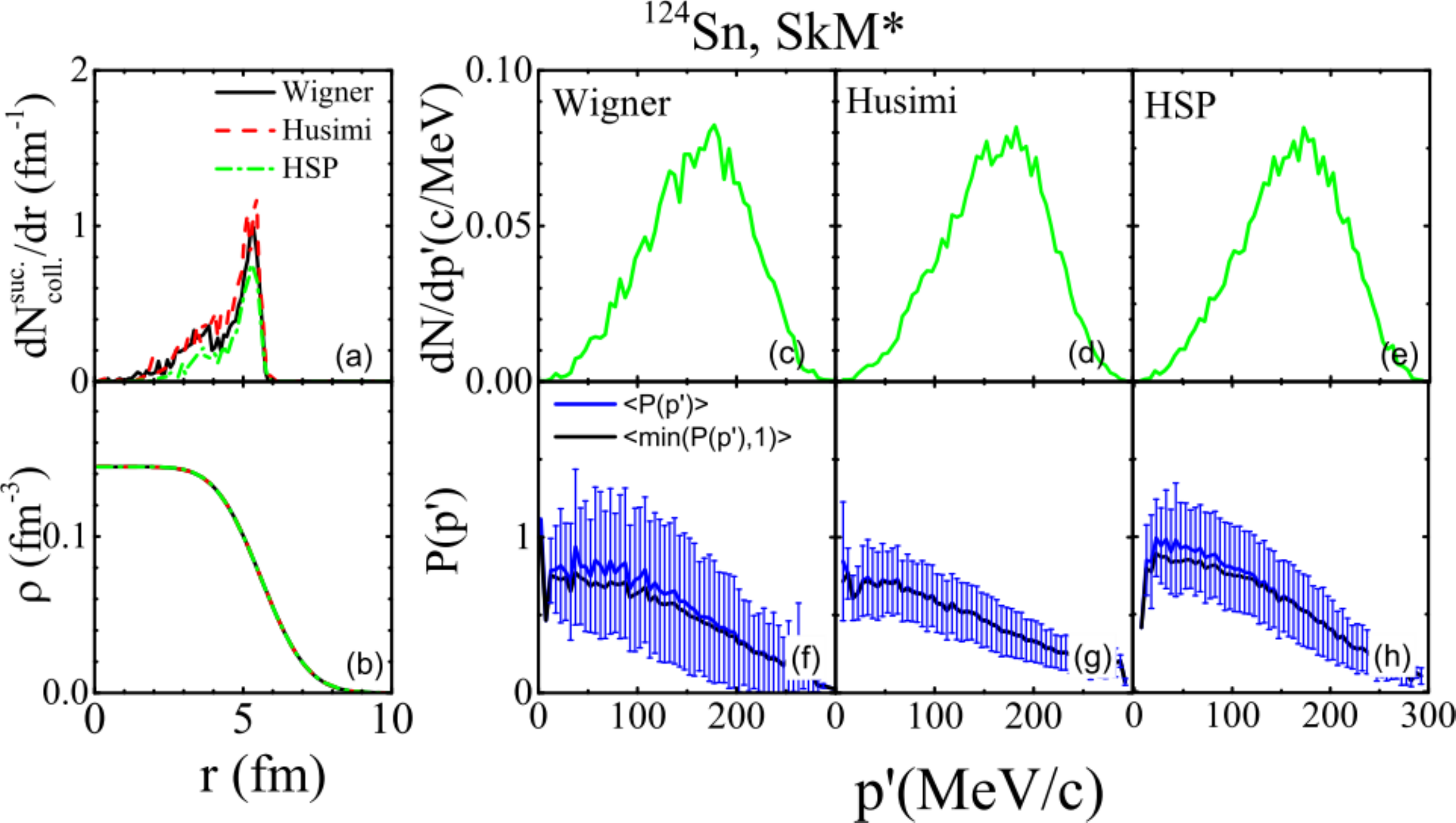}
\setlength{\abovecaptionskip}{0pt}
\vspace{0.0em}
\caption{(Color online) Panel (a): the successful collision number as a function of radial distance for $^{124}$Sn, (b) density distribution of $^{124}$Sn. Panels (c), (d) and (e), momentum distribution of scattered nucleons; (f), (g), and (h): occupation probability for different Pauli blocking algorithms, i.e., PB-Wigner, PB-Husimi, PB-HSP (see the text for details).}
\setlength{\belowcaptionskip}{0pt}
\label{fig:nucleus-coll-coordi}
\end{figure}

Fig.~\ref{fig:nucleus-coll-coordi} (a) shows the number of successful $NN$ collision as a function of radial distance $r$, where the collisions occurs, in $^{124}\mathrm{Sn}$. The lines with different colors represent the results obtained with the PB-Wigner (black line), PB-Husimi (red line), and PB-HSP (green line), respectively. Similar to the calculations in nuclear matter, the numbers of successful collisions are obtained at 1 fm/$c$ and the values of successful collision rates are about 0.9-1.6 $c$/fm. Ideally, the successful collisions rate of initial nucleus should be zero, but it happens owing to the defect of Pauli blocking algorithms in the QMD models. Among the three kinds of Pauli blocking, the calculation with PB-HSP gives the smallest successful collision rate.

In addition, most of the successful $NN$ collisions occur around $r$=5.2 fm, which is close to the surface of nucleus as found in the density distribution plots in panel (b). There were some \textit{ad hoc} methods to overcome this defect in Refs.~\cite{Aichelin1985PRC,Cozma2018EPJA}, but a consistent improvement of Pauli blocking near the surface of nucleus or reaction system is still a theoretical challenge.

The panels (c), (d) and (e) show $dN/dp'$ of the final state of $NN$ collisions in different Pauli blocking algorithms, i.e., PB-Wigner, PB-Husimi, and PB-HSP, respectively, and it clearly illustrates the importance of Pauli blocking within the Fermi momentum. The mean values of occupation probability $\langle P(p')\rangle$ and mean actual occupation probability $\langle$min$(P(p'),1) \rangle$ at 1 fm/$c$ are shown by the blue and black curves in panel (f)-(h), respectively. Similar to the finding in the nuclear matter, there is also $\langle\min(P(p'),1) \rangle \leq \langle P(p')\rangle$ in finite nucleus. The standard deviation of the occupation probability $P(p')$ is indicated by blue error bars, and it shows that the PB-Wigner has the largest fluctuation among the three Pauli blocking algorithms. The values obtained with PB-HSP are closer to 1 at low momentum than other two algorithms.

\subsection{Pauli blocking on stopping power in heavy ion collisions}
\label{HIC}

The nuclear stopping governs the amount of dissipated energy under the competition between the mean field potential and nucleon-nucleon collisions, and can be measured by the ratio between transverse and longitudinal components of kinematical observables~\cite{Reisdorf2004PRL,Lehaut2010PRL,Lopez2014PRC}. For example, the ratio of the variances of the transverse rapidity distribution to that of the longitudinal rapidity distributions of emitted particles, which is named as $vartl$~\cite{Reisdorf2004PRL},
\begin{equation}\label{vartl}
vartl=\frac{\langle y_{t}^{2}\rangle}{\langle y_{z}^{2}\rangle}.
\end{equation}
The energy-based isotropy ratio $R_E$~\cite{Lehaut2010PRL} and
the momentum-based isotropy ratio $R_{p}$~\cite{Lopez2014PRC} are also used in experiments to measure the stopping power.
Those quantities measure the transfer of momentum from entrance direction to transverse direction, and thus, they are closely related to the successful $NN$ collision rate. So, a study of nuclear stopping observable can provide an insight on the in-medium $NN$ cross sections and Pauli blocking effects.

\begin{figure}[htbp]
\centering
\includegraphics[angle=0,scale=0.30]{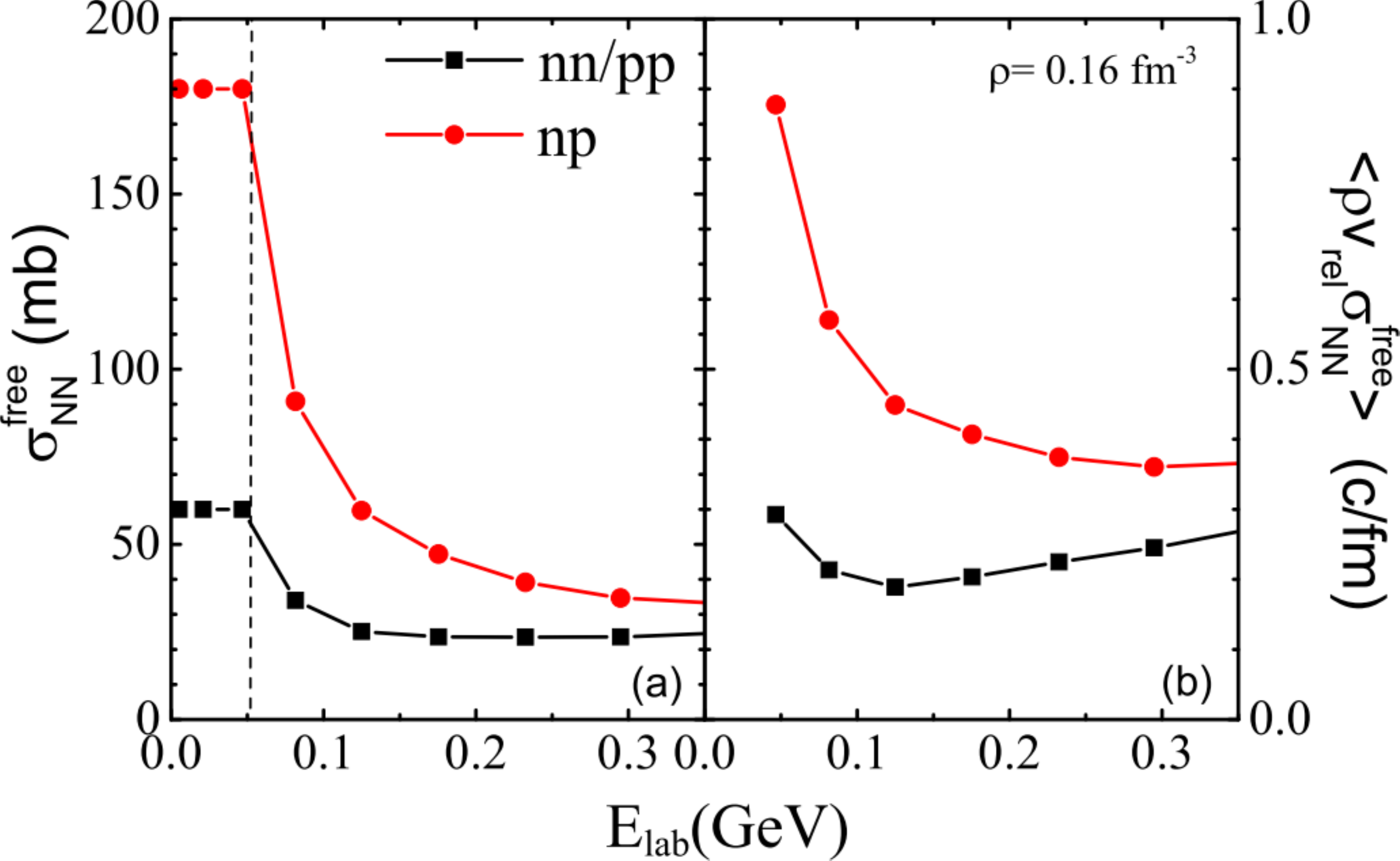}
\setlength{\abovecaptionskip}{0pt}
\vspace{0.2em}
\caption{(Color online). Panel (a), the cross section of nn/pp and np used in ImQMD model; panel (b), the mean attempted collision rate obtained with the free $NN$ cross sections in nuclear matter.}
\setlength{\belowcaptionskip}{0pt}
\label{fig:xs-cugnon-rvs}
\end{figure}

Before discussing the stopping power in HICs, we first present the free $NN$ cross sections~\cite{Cugnon1996NIM} used in the ImQMD calculations and the corresponding mean attempted collision rate in Fig.~\ref{fig:xs-cugnon-rvs} (a) and (b). As shown in panel (a), both the cross sections of nn/pp (black line) and np (red line) decrease with the beam energy. More details, in the ImQMD model, we also set $\sigma^{free}_{nn/pp}=60$ mb and $\sigma^{free}_{np}=180$ mb at $p_{lab}<0.3$ GeV/$c$ (or $E_{lab}<50$ MeV) to avoid the spurious low energy $NN$ collisions in nuclear medium. Fig.~\ref{fig:xs-cugnon-rvs} (b) shows that the mean attempted collision rate by nucleons, i.e., $\langle \rho v_{rel}\sigma^{free}_{NN} \rangle$, for nn/pp and np, in the uniform nuclear matter with $\rho=0.16$ fm$^{-3}$. The mean attempted collision rate for np decreases with the beam energy increasing, while the mean attempted collision rate for nn/pp first decreases and then increases at $E_{lab}>\sim 100$ MeV. It is different with the behaviors in Fig.~\ref{fig:box-pbk-anal} where a constant $NN$ cross section $\sigma=40$ mb is used. Thus, the stopping power may weakly depend on the beam energy or decrease with the beam energy increasing if $\sigma^{free}_{nn/pp,np}$ is used and Pauli blocking is switched off in the model.

\begin{figure}[htbp]
\centering
\includegraphics[angle=0,scale=0.3]{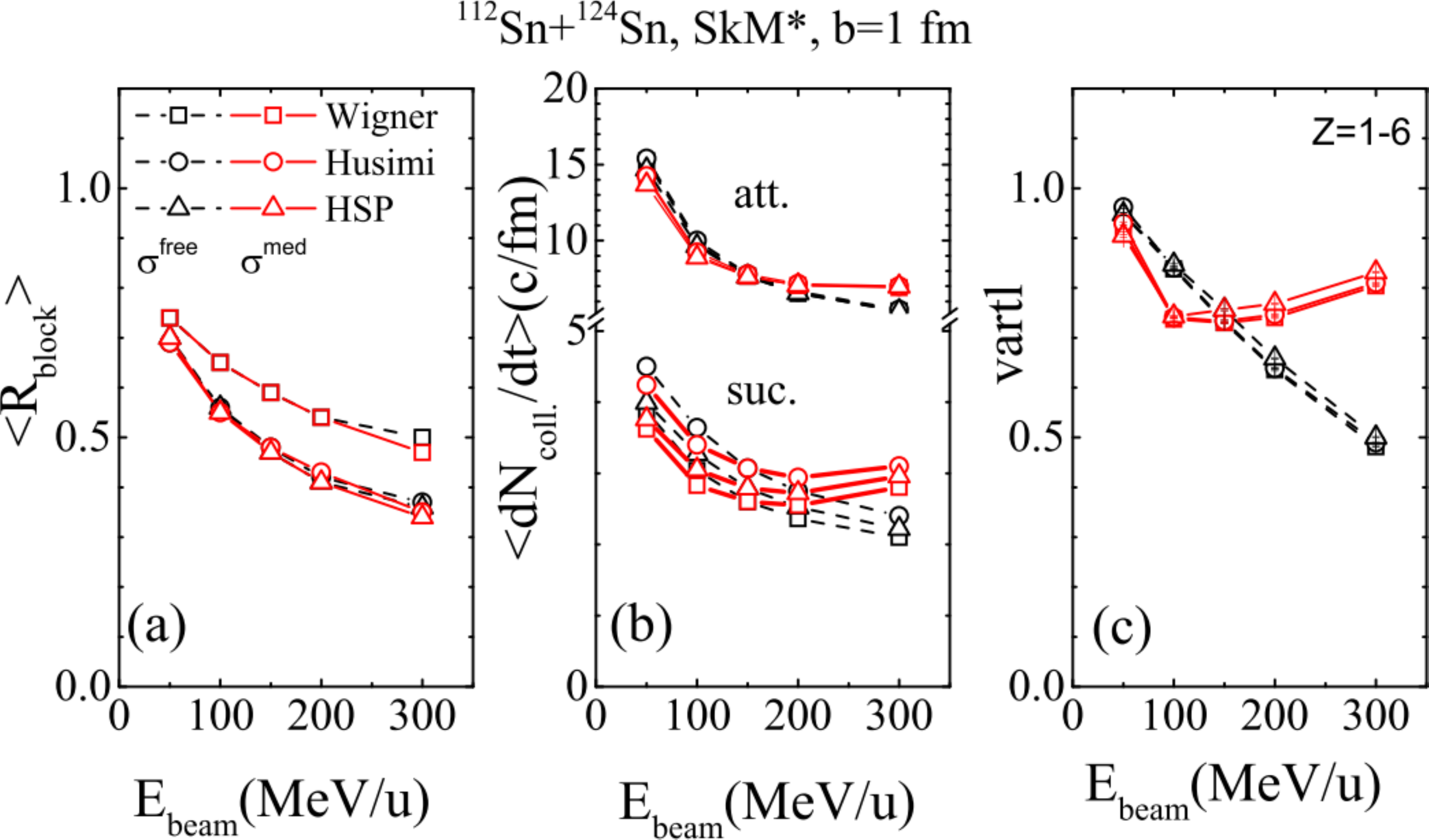}
\setlength{\abovecaptionskip}{0pt}
\vspace{0.2em}
\caption{(Color online). The beam energy dependence of $\langle R_{block}\rangle$ (a), of the mean attempted and successful collision rates (b), and of $vartl$ (c). The different symbols are for different Pauli blocking, and different colors are for different in-medium cross sections.}
\setlength{\belowcaptionskip}{0pt}
\label{fig:reaction-stopping}
\end{figure}

However, the Pauli blocking effect is indispensable for low-intermediate energy heavy ion collisions. In Fig.~\ref{fig:reaction-stopping} (a), the average blocking ratios $\langle R_{block}\rangle$ over the time interval 0-400 fm/$c$ as a function of beam energy for $^{112}Sn+^{124}Sn$ at b=1 fm are presented. The black lines with different symbols are the results obtained with PB-Wigner (squares), PB-Husimi (circles), and PB-HSP (triangles), in the case of $\sigma^{free}$ is adopted in the ImQMD calculations. The values of $\langle R_{block}\rangle$ obtained with PB-Husimi and PB-HSP are almost same, but the PB-Wigner results in the largest $\langle R_{block}\rangle$ among three kinds of Pauli blocking algorithms. It is opposite to the finding in the nuclear matter, and can be understood from the reaction dynamics. At early stage of reaction, the weakest Pauli blocking algorithm observed in the nuclear matter calculations, i.e., PB-Wigner, results in more $NN$ collisions than that for PB-Husim or PB-HSP. More $NN$ collisions provide a larger repulsion for nucleons during the compressed stage, and make the system expand to larger momentum space than that with less $NN$ collisions. Thus, the successful collision rates for PB-Wigner become smaller than that for PB-Husimi or PB-HSP after the compression stage. Thus, the largest $\langle R_{block}\rangle=\langle 1-\frac{dN^{suc.}_{coll.}}{dt}/\frac{dN^{att.}_{coll.}}{dt}\rangle$ values for PB-Wigner are obtained by averaging over the time interval 0-400 fm/$c$. This effect becomes obvious at high beam energy where the $NN$ collisions are more frequent than low beam energies.

To explore the influence of different algorithms of Pauli blocking on the stopping power, we present the averaged attempted and successful collision rates over time interval 0-400 fm/$c$ as a function of beam energy in Fig.~\ref{fig:reaction-stopping} (b) and $vartl$ in panel (c). The values of successful collision rates are in the range of 2-5 $c$/fm, and the different Pauli blocking algorithms lead to about 11-15\% difference on the successful collision rate. The averaged attempted collision rates and the averaged successful collision rates decrease with the beam energy increasing.

The stopping power $vartl$ is calculated from the rapidity distribution of charged particles $Z=1-6$ weighted by their charge number to weaken the defect on the cluster formation mechanism at low-intermediate energy heavy ion collisions. The values of $vartl$ weakly depend on the algorithms of Pauli blocking we used owing to the small number of successful collision rate, as shown in panel (b). Furthermore, the calculations with $\sigma^{free}$ predict that the values of $vartl$ decrease with the beam energy increasing. This behavior is opposite to the observation in experiments\cite{Reisdorf2004PRL} where the $vartl$ increases with the beam energy increasing for Au+Au at $E_{beam}>100$ MeV/u. This difference illustrates that the in-medium correction on the $NN$ cross sections is needed.

\subsection{In-medium $NN$ cross sections on stopping power}
\label{xs}

To test the effect of in-medium $NN$ cross sections on stopping power, we simply take the $\sigma^{med}_{QMD}$ as $\sigma^{med}_{QMD}=(1+\eta(E_{beam})\rho/\rho_0)\sigma^{free}$ in the code. For $E_{beam}\le 100$ MeV, $\eta$ is set as -0.2. At $E_{beam}=150, 200, 300$ MeV, $\eta=0.0, 0.2, 0.8$, respectively. The red lines with symbols in panels (a), (b) and (c) are the results obtained with $\sigma^{med}_{QMD}$. As illustrated in Fig.~\ref{fig:reaction-stopping} (a), the values of $\langle R_{block}\rangle$ weakly depend on the correction of in-medium $NN$ cross sections. But, as shown in panel (b), the in-medium $NN$ cross sections reduce the attempted collision and successful collision rates by $\sim$ 7\% at $E_{beam}\le 100$ MeV and enhance the attempted and successful collision rates by $>7$\% at $E_{beam}>100$ MeV. Consequently, as in panel (c), the values of $vartl$ are influenced and they are suppressed at $E_{beam}\le 100$ MeV/u, and enhanced at $E_{beam}>100$ MeV/u.

However, one should bear in mind that a reliable extraction of in-medium $NN$ cross section by comparing HIC data to transport model calculations requires an accuracy method to treat the Pauli blocking in the simulation of HICs. The improvement of Pauli blocking is still on the way.


\section{Summary}
\label{sec:summary}
In summary, we first evaluate the different Pauli blocking algorithms in the nuclear matter in cascade mode, i.e., only with $NN$ collisions. Our calculations show that averaged occupation probability obtained with PB-Husimi and PB-HSP are closer to the analytical values than that with PB-Wigner, but all the three algorithms used in the QMD codes underestimate the Pauli blocking ratio by 13-25\% at $T\le 10$ MeV. This underestimation may leads to the extracted in-medium $NN$ cross sections from QMD type models are 37-50\% smaller than the true in-medium $NN$ cross sections, and $\sigma^{med}=2\sim2.7\sigma^{med}_{QMD}$.

Furthermore, we analyzed the different Pauli blocking algorithms in the finite nucleus and heavy ion collisions, in which both mean field potential and $NN$ collisions are included. For the finite nucleus, the Pauli blocking ratios are in the range of 69-83\% for different Pauli blocking algorithms owing to the defects of Pauli blocking in QMD model. The spurious $NN$ collision mainly occurs around the surface of finite nucleus. There have been some effort to improve the Pauli blocking, especially near the surface of nucleus or the reaction systems, but a consistent treatment is still a big challenge for many-body transport theory.

By using the algorithms of Pauli blocking in the current market, the influences of different Pauli blocking algorithms on the excitation function of stopping power in HICs are discussed, and the uncertainties of stopping power with different Pauli blocking are less than 5\%.
If one would like to produce the behaviors of the $vartl$ increase with the beam energy, a strong enhancement of in-medium $NN$ cross sections is needed. For obtaining the true values of in-medium $NN$ cross sections by comparing the heavy ion collision data to the transport model calculations, a refined Pauli blocking algorithm is urged in future.

\section*{Acknowledgements}
The authors thank the helpful discussions with Dr. Yongjia Wang, Dr. D. Cozma on the Pauli blocking algorithms. This work was partly inspired by the transport code comparison project, and it was supported by the National Natural Science Foundation of China Nos. 11875323, 11705163, 11790320, 11790323, and 11961141003, the National Key R\&D Program of China under Grant No. 2018YFA0404404, the Continuous Basic Scientific Research Project (No. WDJC-2019-13, BJ20002501) and the funding of China Institute of Atomic Energy.

\end{document}